# Towards a General Definition of Biometric Systems

Markus SCHATTEN[1], Miroslav BAČA[1] and Mirko ČUBRILO[1]

[1] Faculty of Organization and Informatics, University of Zagreb
Pavlinska 2, 42000 Varaždin, Croatia
*{markus.schatten, miroslav.baca, mirko.cubrilo}@foi.hr*

**Abstract**
A foundation for closing the gap between biometrics in the narrower and the broader perspective is presented trough a conceptualization of biometric systems in both perspectives. A clear distinction between verification, identification and classification systems is made as well as shown that there are additional classes of biometric systems. In the end a Unified Modeling Language model is developed showing the connections between the two perspectives.

## 1. Introduction

The term biometrics comming from ancient greek words $\beta\iota o\varsigma$ (bios) for life and $\mu\varepsilon\tau\rho o\nu$ (metron) for measure is often used in different contexts to denote different meanings. At the same time there are very similar and often synonimic terms in use like biometry, biological statistics, biostatistics, behaviometrics etc. The main aim of this paper is to show the connection between these various views of biometrics as well as to continue our research on the essence of biometric systems.

In [2] we showed how to apply a system theory approach to the general biometric identification system developed by [8] in order to extend it to be aplicable to unimodal as well as multimodal biometric identification, verification and classification systems in the narrower (security) perspective of biometrics. The developed system model is partialy presented on figure 1.

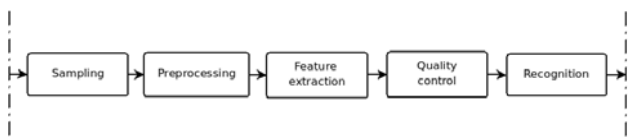

Fig. 1 Pseudo system diagram of the developed model.

In [3] we argued that there is a need for an open biometrics ontology that was afterwards partially build in [1] and [7]. During the development of this ontology crucial concepts like biometric system, model, method, sample, characteristic, feature, extracted structure as well as others were defined. We also developed a full taxonomy of biometric methods in the narrower perspective in [6] that contributed to a unique framework for communication.

All this previous research showed that there is confusion when talking about different types or classes of biometric systems. Most contemporary literature only makes distinction between verification and identification systems but some of our research showed that there are more different classes like simple classification systems that seem to be a generalization of verification as well as identification systems [7]. As we shall show in our following reasoning by taking the input and output sets of the different processes in biometric systems that define biometric methods into consideration, as well as mappings between them a concise conceptualization emerges that seems to applicable to any biometric system.

## 2. Basic Definitions in Biometrics

In order to reason about biometrics we need to introduce some basic definitions of concepts used in this paper. These definitions were crutial to the development of a selected biometrics segments ontology as well as an taxonomy of biometric methods.

First of all, we can approach biometrics in a broader and in a narrower perspective as indicated before. In the broader perspective biometrics is the statistical research on biological phenomenae; it is the use of mathematics and statistics in understanding living beeings [4]. In the narrower perspective we can define biometrics as the research of possibilities to recognize persons on behalf of their physical and/or behavioral (psychological) characteristics. We shall approach biometrics in the broader perspective in this paper.

A **biometric characteristic** is a biological phenomenon's physical or behavioral characteristic that can be used in order to recognize the phenomenon. In the narrower perspective of biometrics **physical characteristics** are genetically implied (possibly environmental influenced) characteristics (like a person's face, iris, retina, finger, vascular structure etc.). **Behavioral or psychological characteristics** are characteristics that one acquires or learns during her life (like a handwritten signature, a person's gait, her typing dynamics or voice characteristics). These definitions are allmost easily translated into the broader perspective of biometrics. Depending on the number of characteristics used for recognition biometric systems can be unimodal (when only one biometric characteristic is used) or multimodal (if more than one characteristic is used).

A **biometric structure** is a special feature of some biometric characteristic that can be used for recognition (for example a biometric structure for the human biometric characteristic finger is the structure of papillary lines and minutiea, for the human biometric charactersitic gait it is the structure of body movements during a humans walk etc.).

The word method comes from the old greek $\mu\varepsilon\theta o\delta o\varsigma$ (methodos) that literally means "way or path of transit" and implies an orderly logical arrangement (usually in steps) to achieve an attended goal [9; pp. 29]. Thus a **biometric method** is a series of steps or activities conducted to process biometric samples of some biometric characteristic usually to find the biometric characteristic's holder (in the narrower perspective) or a special feature of the biometric sample (in the broader perspective).

A model is a (not neccesarily exact) image of some system. It's main purpose is to facilitate the aquiring of information about the original system [5; pp. 249]. A **biometric model** is thus a sample of a biometric system that facilitates the aquiring of information about the system itself as well as information about biometric characteristics. In [2] and [7] we showed that biometric models consist of biometric methods for preprocessing and feature extraction, quality control as well as recognition.

A sample is a measured quantity or set of quantities of some phenomenae in time and/or space. Thus a **biometric sample** represents a measured quantity or set of quantities of a biological phenomenae [7].

A **biometric template or extracted structure** is a quantity or set of quantities aquired by a conscious application of a biometric feature extraction or preprocessing method on a biometric sample. These templates are usually stored in a biometric database and used for reference during recognition, training or enrollment of a biometric system.

## 3. Conceptualizing Input and Output Mappings

Having the basic concepts defined we can formalize the domain using the following seven sets: (1) $S_m$ as the set of all biometric samples, (2) $S_p$ as the set of all preprocessed samples, (3) $T$ as the set of all biometric templates or extracted structures, (4) $T_q$ as a subset of $T$ representing all extracted structures that are suitable for recognition after quality control, (5) $B_p$ as the set of all biological phenomenas (in the broader perspective) or all persons (in the narrower one) represented by biometric structures on behalf of which recognition is made possible, (6) $B_e$ as a subset of $B_p$ of all biological phenomenas that are enrolled, and (7) $C$ as the set of all recognition classes.

Using these sets we can formalize the classes of biometric methods shown on figure 1. The sampling process, the preprocessing, the feature extraction process, the quality control process as well as the recognition process are described using the mappings shown in the following set of equations respectively:

$$S : B_p \longrightarrow S_m$$
$$P : S_m \longrightarrow S_p$$
$$F : S_p \longrightarrow T$$
$$Q : T \longrightarrow T_q$$
$$R : T_q \longrightarrow C$$

Figure 2 shows the mapping of the sampling process. One can observe that every value from $S_m$ has its argument in $B_p$. Arguments from $B_p$ can have 0 or more values in $S_m$. This can be explained easily because there is a high probability that not every biological phenomenon will be sampled by one biometric system, but every biometric sample is a sample of a real biological phenomenas.\footnote.[1] There is also a considerable probability that a group of biological phenomenas will yield the same sample which is depending on the quality of the sampling technology.

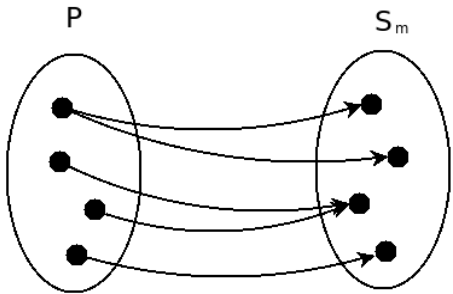

Fig. 2  Mapping of the sampling process.

In multimodal systems one sample can be made on behalf of more than one feature, what would yield a different figure than the one above. But, since every sample is made on behalf of exactly $\mu$ features (where $\mu$ is the number of characteristics used in the multimodal system) we can consider the tuple $(p_1, p_2, ..., p_\mu)$ (where $p_1, p_2, ..., p_\mu$ are partial features that are being sampled) to be only one feature. The mapping would thus have $\mu$ arguments and the figure wouldn't change, or likewise the set $B_p$ would consist of tuples.

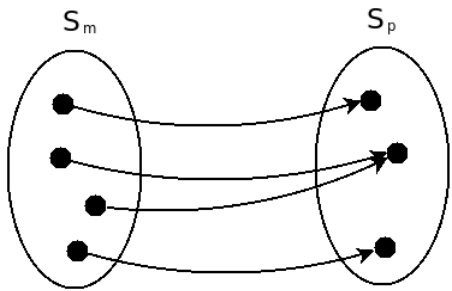

Fig. 3  Mapping of the preprocessing process.

Figure 3 shows the mapping of the preprocessing process which happens to be a function (since every element from the domain $S_m$ is associated with some element in the co-domain $S_p$). The function is surjective but not necessarily injective since some samples can yield the same preprocessed sample even if they are distinct.

---

[1]　Presuming that there are no fake biometric samples in $S_m$

Figure 4 shows the mapping of the feature extraction process which also happens to be a function (since every element from the domain $S_p$ is associated with some element in the co-domain $T$). The function is likewise surjective and likewise not necessarily injective since some preprocessed samples can yield the same extracted structure even if they are distinct.

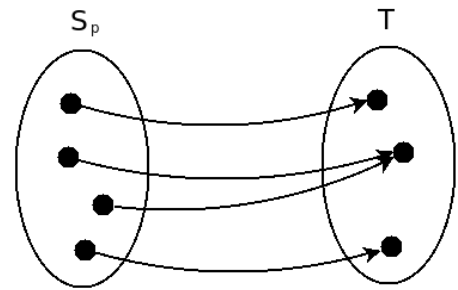

Fig. 4  Mapping of the feature extraction process.

There is another possibility in multimodal systems, when samples aren't multimodal, but structures are extracted from multiple samples. As in the case of the sampling process mentioned before we can consider the elements in $S_p$ to be tuples of samples $(s_1, s_2, ..., s_\mu)$ (where $\mu$ is the is the number of characteristics used in the multimodal system) that are used to extract a single structure.

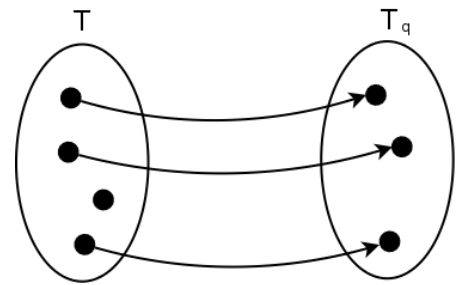

Fig. 5  Mapping of the quality control process.

The mapping of the quality control process is shown on figure 5. Every value from $T_q$ has its argument in $T$ but the opposite does not necessarily hold true since some extracted structures do not pass the quality test and are abandoned. Thus, every argument from $T$ has 0 or 1 values in $T_q$ and the values are unique.

Likewise figure 6 shows the mapping of the recognition process that is similar to the previous one. Again, every value from $C$ has its argument in $T_q$ but the opposite is not necessarily true since some structures that passed the quality test cannot be recognized and classified into one of the classes for recognition in $C$. We could define a set $C^+ \simeq C \cup \{\nu\}$ where $\nu$ is the class for all unrecognized structures but we left this part out due to concept consistency and simplicity. Thus, every argument in $T_q$ has 0 or 1 image in $C$ whereby the images are not necessarily unique.

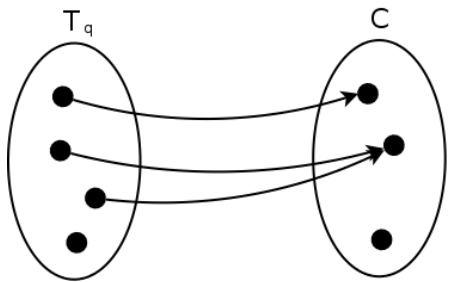

Fig. 6  Mapping of the recognition process.

Special cases of the recognition process mapping include the case when $C \simeq B_e$ and the case when $R$ is a mapping of two variables. In the former case we have the mapping $R_i : T_q \longrightarrow B_e$ that represents an actively functioning biometric identification system. In the letter case we have the mapping $R_v : T_q^+ \longrightarrow B_e$ (whereby elements of $T_q^+$ are tuples $(\tau, \pi)$ where $\tau \in T_q$ and $\pi \in B_p$) that represent an actively functioning biometric verification system.

## 4. Conceptualizing Mapping Cardinalities of the Recognition Process

If we consider the mapping $R : T_q \longrightarrow C$ of the recognition process and presume that the biometric system is active (thereby eliminating passive periods) we can observe the following situations:

- If one tuple of one person information and one extracted structure are mapped to exactly one class and $C \simeq B_e$ then the system is a biometric verification system in normal (active) functioning. We denote this mapping with $1 : 1, C \simeq B_e$.

- If one extracted structure is being mapped to one of $n$ classes where $n$ is the cardinality of set $C$ then the system is a biometric classification system in normal (active) functioning. We denote this mapping with $1 : n, c(C) = n$.

- If one extracted structure is being mapped to one of $n'$ classes where $n'$ is the cardinality of set $B_e$ and $C \simeq B_e$ then the system is a biometric identification system in normal active functioning. We denote this mapping with $1 : n', C(B_e) = n'$.

- If $m$ tuples of person information and extracted structures (where $m$ is the cardinality of set $T_q^+$) are being mapped individually into exactly one class and when $C \simeq B_e$ then the system is a biometric verification system during training. We denote this mapping with $m : 1, c(T_q^+) = m$.

- If $m$ extracted structures are being mapped individually into one of $n$ classes then the system is a biometric classification system during training. We denote this mapping with $m : n, c(T_q) = m, c(C) = n$.

- If $m$ extracted structures are being mapped individually into one of $n'$ classes and $C \simeq B_e$ then the system is a biometric identification system during training. We denote this mapping with $m : n', c(T_q) = m, c(B_e) = n'$.

- If $n'$ groups of $k$ tuples consisting of person information and extracted structure are being mapped into exactly one of $n'$ classes (whereby $c(T_q^+)/k = c(B_e) = n'$, and $k$ is the number of extracted structures per person)[2] the system is a biometric verification system during enrollment. We denote this mapping with $n' \cdot k : 1, c(T_q)/k = c(B_e) = n'$.

- If $n$ groups of $k$ extracted structures are being mapped into one of $n$ classes (whereby $c(T_q)/k = c(C) = n$, and $k$ is the number of extracted structures per class) the system is a biometric classification system during enrollment. We denote this mapping with $n \cdot k : n, c(T_q)/k = c(C) = n$.

---

2  Usually a standard number of samples is used for enrollment but $k$ can be variable due to lack of such standard or due to eliminated samples during other processes of the biometric system.

- If $n'$ groups of $k$ extracted structures are being mapped into one of $n'$ classes (whereby $c(T_q)/k = c(B_e) = n'$, and $k$ is the number of extracted structures per person) the system is a biometric identification system during enrollment. We denote this mapping with $n' \cdot k : n', c(T_q)/k = c(B_e) = n'$.

From this reasoning we can conclude that biometric verification and identification systems are only special cases of biometric classification systems when the number of classes into which extracted structures are mapped into are equivalent to the set of biological phenomenas (or persons in the narrower sense) that are enrolled. Further we can observe three distinct situations in biometric systems recognition process cardinalities defined in equation $\Upsilon : \Omega$ where $\Upsilon$ is the number of extracted structures (or tuples in the case of verification systems) on the input to the recognition process, and $\Omega$ the number of classes (outputs) into which the inputs are being mapped.

1. In the case when $\Upsilon = 1$ and $\Omega = c(C)$ the biometric system is in **normal** (everyday) use.
2. In the case when $\Upsilon = c(T_q)$ and $\Omega = c(C)$ the biometric system is in the **training** phase.
3. In the case when $\Upsilon = c(C) \cdot k$ and $\Omega = c(C)$ the biometric system is in the **enrollment** phase (whereby $k$ is an positive integer possibly inside an interval, $k \in \langle min, max \rangle$).

## 5. Conceptualizing Relations Between the Defined Sets

Figure 7 shows the UML (*Unified Modeling Language*) class diagram of the defined concepts that gives us even deeper insight of the domain being conceptualized. Every class applies for some of the previously defined sets. The class **Phenomenon** applies to the set $B_p$ of all biological phenomenas. As the diagram shows there is a special subset defined by the class **Person**. Every **Person** instance is an instance of **Phenomenon**. There are also two other special subsets of $B_p$ denoted by the set $B_e$ of all enrolled phenomenas or persons in the narrower sense of biometrics. Thus every instance of **Enrolled phenomenon** is an instance of **Phenomenon**, every instance of **Enrolled person** is an instance of **Person**, as well as every **Enrolled person** is an instance of **Enrolled phenomenon**.

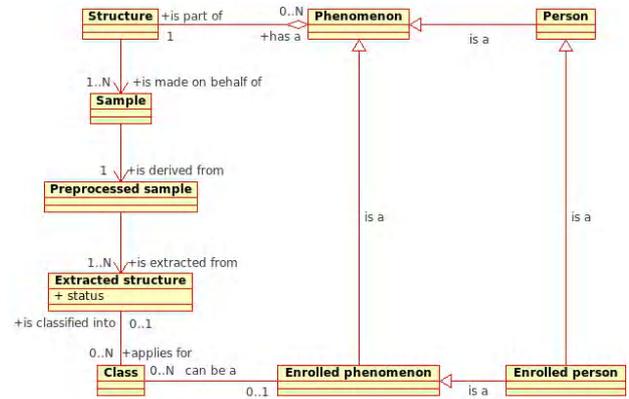

Fig. 7  UML Class Diagram of the Defined Concepts.

As shown in the diagram any **Phenomena** can consist of zero or more biometric **Structure** instances, while a biometric **Structure** is part of exactly one biological **Phenomena**. Following the process flow from figure 1 we can observe that every **Sample** instance is made on behalf of one or more **Structure** instances[3] and thus the set $S_m$ is represented with the class **Sample**. The case that a sample is made on behalf of more biometric structures applies only to multimodal systems where multiple biometric structures are sampled into exactly one sample.

Every instance of **Preprocessed sample** instance is derived from exactly one **Sample** instance where the class **Preprocessed sample** corresponds to the set $S_p$. Further on every **Extracted structure** instance is extracted from one or more **Preprocessed sample** instances. The class **Extracted structure** represents the two sets concerning biometric templates or extracted structures $T$ and $T_q$ depending on the value of an instance's status attribute. If the value is **untested** or **failed** the instance belongs into set $T$ (since the instance hasn't been tested for quality or it hasn't pass the quality test). In the opposite case when the value is

---

3    Presuming again that there are only real biometric samples in $S_m$

**passed** the instance belongs into the $T_q$ set since it has been tested for quality and passed the test. The enumeration holding the values of the status attribute has been left out form the diagram for the sake of simplicity. The case when an extracted structure is extracted from more biometric samples applies only to multimodal biometric systems that extract features on behalf of more biometric samples, whilst the case when on extracted structure is extracted from only one sample applies to unimodal biometric systems.

Every **Extracted structure** can be classified into zero or more instances of **Class** whilst every **Class** instance applies to zero or more instances of **Extracted structure**. The **Class** class represents the set $C$ as it is obvious from our previous reasoning. There is correspondence between the **Class** class and the **Enrolled phenomenon** class depending on the purpose of the system as argued before.

From this reasoning we can conclude that the classes **Structure**, **Sample**, **Preprocessed sample**, **Extracted structure** and **Class** apply to both biometrics in the narrower and the broader perspective. If the connected classes are **Phenomenon** and **Enrolled phenomenon** we are talking about the broader perspective of biometrics. In the other case when the connected classes are **Person** and **Enrolled person** the narrower perspective comes into play. Since **Person** is a special case of **Phenomenon** and **Enrolled person** is a special case of **Enrolled phenomenon** the narrower perspective of biometrics is only a special case of the broader one.

## 6. Conceptualizing Relations Between the Defined Sets

In this paper we showed a simple conceptualization of biometric systems. If one considers a general biometric system consisting of a series of processes she can observe the input and output sets of any given process. By mapping these sets in a sequence of events one can observe their features. The recognition process is of special interest since the special cases of the possible mappings define the three types of biometric systems (classification, verification, identification) as well as the three possible processing conditions (everyday use, training, enrollment).

As we showed, biometric verification and identification systems are only special cases of biometric classification systems where the number of classes into which samples are classified into is equivalent to the number of enrolled biological phenomenas (in the broader sense of biometrics) or the number of enrolled persons (in the narrower perspective).

We argued that biometrics in the narrower and in the broader perspective have a lot in common especially when talking about data and data manipulation techniques. Biometrics in the narrower perspective is and remains a special case of biometrics in the broader perspective. Thus this conceptualization presents a clear framework for communication on any biometric system topic.

The only thing that seems to be the difference is the semantic context in which the same methods are used. So we ask our self, why making a difference? The developed UML model merges the two perspectives by stating that biometrics in information sciences and information system security specialization of biometrics in mathematics, statistics and biology. The narrower perspective heavily depends on theories from the broader one, but insights from information system's security biometrics are of course usefull in the biology, mathematics and statistics perspective especially when talking about system planning and implementation.

If we add this conceptualization to our previously developed open ontology of chosen parts of biometrics, as well as to the developed systematization and taxonomy of biometric methods, characteristics, features, models and systems we get an even clearer framework for communicating about biometrics that puts our research into a broader perspective. Future research shall yield an open ontology of biometrics in the broader perspective.

**Acknowledgments**

Results presented in this paper came from the scientific project "Methodology of biometrics characteristics evaluation" (No. 016-0161199-1721) supported by Ministry of Science Education and Sports Republic of Croatia.

**References**
[1] M. Bača, M. Schatten, and B. Golenja, "Modeling Biometrics Systems in UML". in **IIS2007 International Conference on Intelligent and Information Systems** Proceedings. 2007, Vol. 18, pp. 23–27.
[2] M. Bača, M. Schatten, and K. Rabuzin, "A Framework for

IJCSI

Systematization and Categorization of Biometrics Methods". in **IIS2006 International Conference on Intelligent and Information Systems** Proceedings. 2006, Vol. 17, pp. 271–278.

[3] M. Bača, M. Schatten, and K. Rabuzin, "Towards an Open Biometrics Ontology", **Journal of Information and Organizational Sciences**, Vol. 31, No. 1, 2007, pp. 1–11.

[4] R. H. Jr. Giles, "Lasting Forests Glossary". Available at http://fwie.fw.vt.edu/rhgiles/appendices/glossb.htm, Accessed: 28th February 2005.

[5] D. Radošević, **Osnove teorije sustava**, Zagreb: Nakladni zavod Matice hrvatske, 2001.

[6] M. Schatten, M. Bača, and K. Rabuzin, "A Taxonomy of Biometric Methods", in **ITI2008 International Conference on Information Technology Interfaces** Proceedings, Cavtat/Dubrovnik: SRCE University Computing Centre 2008; 389–393.

[7] M. Schatten, "Zasnivanje otvorene ontologije odabranih segmenata biometrijske znanosti" M.S. Thesis, Faculty of Organization and Informatics, University of Zagreb, Varaždin, Croatia, 2008.

[8] J. L. Wayman, "Generalized Biometric Identification System Model", in **National Biometric Test Center Collected Works 1997. - 2000**. San Jose: San Jose State University. 2000, pp. 25–31.

[9] M. Žugaj, K. Dumičić and V. Dušak, **Temelji znanstvenoistraživačkog rada. Metodologija i metodika**, Varaždin: TIVA & Faculty of Organization and Informatics, 2006.

**M. Schatten** received his bachelors degree in Information systems (2005), and his masters degree in Information Sciences (2008) both on the faculty of Organization and Informatics, University of Zagreb where he is currently a teaching and research assistent. He is a member of the Central European Conference on Intelligent and Information Systems organizing comitee. He is a researcher in the Biometrics center in Varaždin, Croatia.

**M. Bača** received his bachelor degree form Faculty of Electrical Engineering in Osijek (1992), second bachelors degree form High Police School in Zagreb (1996), MSc degree form Faculty of Organization and Informatics, Varaždin (1999), PhD degree from Faculty of Organization and Informatics, Varaždin (2003). He was an Assistant professor, University of Zagreb, Faculty of Organization and Informatics (2004-2007), and is currently an Associated professor, University of Zagreb, Faculty of Organization and Informatics. He is a member of various professional societies and head of the Central European Conference on Intelligent and Information Systems organizing comitee. He is also the head of the Biometrics center in Varaždin, Croatia. He lead 2 scientific projects granted by the Ministry of Science, Education and Sports of Croatia.

**M. Čubrilo** received his bachelors (1979) and masters (1984) degree from the Faculty of Natural Sciences and Mathematics, University of Zagreb. He received his PhD degree from the Faculty of Electrotechnics, University of Zagreb (1992). He is currently a full professor at the Faculty of Organization and Informatics, University of Zagreb. He was main editor of the Journal of Information and Organization Sciences. He was leader of 2 scientific projects granted by the Ministry of Science, Education and Sports of Croatia.

IJCSI